\RequirePackage{rotating}
\RequirePackage{fix-cm}

\documentclass{article}
\usepackage{arxiv}

\usepackage{subcaption}
\usepackage{graphicx}
\usepackage{cite}
\usepackage{amsmath,amssymb,amsfonts}
\usepackage{algorithm}
\usepackage{algpseudocode}
\usepackage{algorithmicx}
\usepackage{graphicx}
\usepackage{textcomp}
\usepackage{xcolor}
\usepackage{paralist}
\usepackage{bm}
\usepackage{adjustbox}

\title{The Cost Advantage of Virtual Machine Migrations: Empirical Insights into Amazon's EC2 Marketspace}

\author{Benedikt Pittl\\
	Faculty of Computer Science\\
	University of Vienna\\
	A-1090 Vienna, Austria\\
	\texttt{benedikt.pittl@univie.ac.at} \\
	\And
	Werner Mach\\
	Faculty of Computer Science\\
	University of Vienna\\
	A-1090 Vienna, Austria\\
	\texttt{werner.mach@univie.ac.at} \\
    \AND
    Erich Schikuta\\
	Faculty of Computer Science\\
	University of Vienna\\
	A-1090 Vienna, Austria\\
	\texttt{erich.schikuta@univie.ac.at} \\
}

\begin{document}

\date{}

\maketitle

%






\begin{abstract}
In recent years, cloud providers have introduced novel approaches for trading virtual machines. For example, Virtustream introduced so-called $\mu$VMs to charge cloud computing resources while other providers such as Google, Microsoft, or Amazon re-invented their marketspaces. Today, the market leader Amazon runs six marketspaces for trading virtual machines. Consumers can purchase bundles of virtual machines, which are called cloud-portfolios, from multiple marketspaces and providers. An industry-relevant field of research is to identify best practices and guidelines on how such optimal portfolios are created.
In the paper at hand, a cost analysis of cloud portfolios is presented. Therefore, pricing data from Amazon was used as well as a real virtual machine utilization dataset from the~\emph{Bitbrains} datacenter. The results show that a cost optimum can only be reached if heterogeneous portfolios are created where virtual machines are purchased from different marketspaces. Additionally, the cost-benefit of migrating virtual machines to different marketplaces during runtime is presented. Such migrations are especially cost-effective for virtual machines of cloud-portfolios which run between $6$ hours and $1$ year. The paper further shows that most of the resources of virtual machines are never utilized by consumers, which represents a significant future potential for cost optimization. For the validation of the results, a second dataset of the Bitbrains datacenter was used, which contains utility data of virtual machines from a different domain of application.
\end{abstract}

\section{Introduction}

Corporations try to establish novel business models to gain shares on the cloud market~\cite{LabesHZ17}. For example, Virtustream introduced $\mu$VMs as a metric to measure and charge computational power while the German Stock Exchange tried to establish a centralized trading platform called~\emph{Deutsche Boerse Cloud Exchange}. This platform was intended as a portal where consumers were able to purchase virtual machines from different cloud providers. However, after a couple of months, the portal closed in 2016. The low number of providers who participated is considered the main reason for the platform's failure~\cite{Herrmann2016}. The market leader, Amazon, continuously updates its platform where consumers can purchase virtual machines. For example, the pricing model of Amazon's spot marketplace was relaunched at the end of 2017. Before the spot marketspace was relaunched, consumers were able to bid for virtual machines. If the bid exceeded the so-called spot market price, which is a dynamic price determined by Amazon, then consumers were able to use the virtual machine. If not, then the virtual machine was interrupted. On the new spot marketspace, consumers need not bid anymore, and the prices for virtual machines are more stable than they were before. However, Amazon can reclaim the resources at any time, which results in interruptions of virtual machines.  Therefore, the prices on the spot marketspace are significantly lower\footnote{https://aws.amazon.com/blogs/compute/new-amazon-ec2-spot-pricing/} than the prices on other marketspaces where such interruptions are not possible. An excellent analysis of the new spot marketspace is given in~\cite{ChhetriLVK18}. In addition to the spot marketspace, Amazon offers further marketspaces:
\begin{inparaenum}[(i)]
	\item Similar to virtual machines from the spot marketspace, virtual machines from the so-called 6-hour spot marketspace and the 1-hour spot marketspace are charged per hour. These virtual machines are not interrupted but come with a fixed, predefined lifetime.
	\item The on-demand marketspace is Amazon's most prominent marketspace where consumers are charged per hour. Virtual machines from the on-demand marketspace are neither interrupted nor do they have a fixed, predefined lifetime.
	\item The reservation marketspace allows consumers to purchase virtual machines from one year to three years by paying a predefined fee. The hourly prices of virtual machines on the reservation marketspace are significantly lower than the prices on the on-demand marketspace. 
\end{inparaenum}


 Usually, institutions run a bundle of virtual machines that have different characteristics. Some of the virtual machines are hosted on private clouds, while the rest have to be hosted on public clouds. For those virtual machines, institutions can choose between different providers and marketspaces. A challenging problem in industry as well as in academia is to find a cost-optimal cloud portfolio. In the paper at hand, we use a real dataset of virtual machine utilization from Bitbrains datacenter\footnote{http://gwa.ewi.tudelft.nl/datasets/gwa-t-12-Bitbrains}. We treat those virtual machines as~\emph{requested} virtual machines for which we form optimal homogeneous and heterogeneous portfolios using the  marketspaces from Amazon. Different pricing models - e.g., fees, hourly pricing - as well as technical constraints such as unpredictable interruptions of virtual machines purchased on Amazon's EC2 spot marketspace have to be considered when creating such cloud-portfolios. To the best of the authors' knowledge, neither the scientific community nor the industry has presented an empirical analysis or a case study that addresses this problem. We see this paper as a first step towards creating cost-efficient cloud portfolios.

The paper is structured as follows: Section~\ref{sec:foundations} summarizes foundations and related work. In section~\ref{sec:problem} we formalize the cloud-portfolio optimization problem. The portfolio creation process is presented in section~\ref{sec:instancetypemapping}. The structure of homogeneous cloud-portfolios is given in section~\ref{sec:homoPort}, followed by an analysis of the structure of heterogeneous portfolios in section~\ref{sec:costAnalysis}. Section~\ref{sec:cost} presents a cost analysis of the created portfolios. We use a second dataset of the Bitbrains datacenter to validate our findings in section~\ref{sec:validation}. The paper closes with a conclusion in section~\ref{sec:conclusion}.

\section{Foundations and Related Work}
\label{sec:foundations}

The scientific community proposed over time different visions of future cloud markets, identifying economic aspects even in Grids~\cite{weishaupl2005business, schikuta2005business,schikuta2008workflow}, the computing paradigm preceding Clouds. 

Trust and security are critical enablers for the digital economy because they foster confidence among participants, ensuring that individuals and businesses are willing to engage in digital transactions and share sensitive data. Without trust~\cite{weishaupl2006gset}, users would hesitate to adopt digital platforms, and without security~\cite{shaaban2019ontology}, the risk of fraud, data breaches, and cyberattacks would undermine the integrity of the ecosystem. 

This section introduces related work as well as virtual machine migration techniques on Amazon.

The idea of dynamic cloud-markets was discussed by several authors, as~\cite{ChichinVK17,PittlHMS17wetice,bonacquisto_strategy_2014,DastjerdiB15,pittl2015negotiation}. All these papers used the notion of cloud-portfolios, but with a focus on the introduction of novel marketspaces, they neglected a detailed analysis of them. An excellent paper on cloud-portfolios by~\cite{IrwinSSS17} states that cloud-portfolios face, similar to financial portfolios, a tradeoff between risk and profit. The authors introduce risk management techniques for reducing interruptions of virtual machines, such as hedging or active trading. However, with the focus on risk management, the creation and analysis of cost-efficient cloud portfolios were neglected. Further, no comprehensive use case was introduced. Based on that work, the authors introduced the project~\emph{ExoSphere} for risk modeling and analysis of cloud-portfolios~\cite{SharmaIS17}. 
Joe Weinman mentioned the importance of cloud-portfolios in several publications, e.g.~\cite{Weinman16a,Weinman15}, without presenting a detailed analysis of cloud-portfolios. Several papers focused on the analysis of Amazon's spot marketspace: The authors of~\cite{bs-1811-12901} evaluated bidding strategies on the spot marketplace. Thereby, the authors assume that consumers can choose between the on-demand marketspace and the spot marketspace. Concrete datasets, as well as other marketspaces were neglected. The authors of~\cite{ChhetriLVK18} investigated the pricing differences of the~\emph{old} spot marketspace - before 2017 - and the~\emph{new} spot marketspace. Cloud portfolios were not considered. Similarly, the authors of~\cite{hamRF18} empirically determined the frequency of interruptions of virtual machines on Amazon's EC2 spot marketspace without considering the creation of cloud-portfolios. The trade-off between investing in private and public clouds can be considered as a typical IT-investment which was, e.g., investigated by~\cite{MullerSZH16} and~\cite{BuhlHPS16}. With a focus on a generic IT-investment framework, cloud portfolios were neglected.

\begin{figure}[ht]
\begin{subfigure}[c]{0.45\textwidth}
	\includegraphics[width=0.99\linewidth]{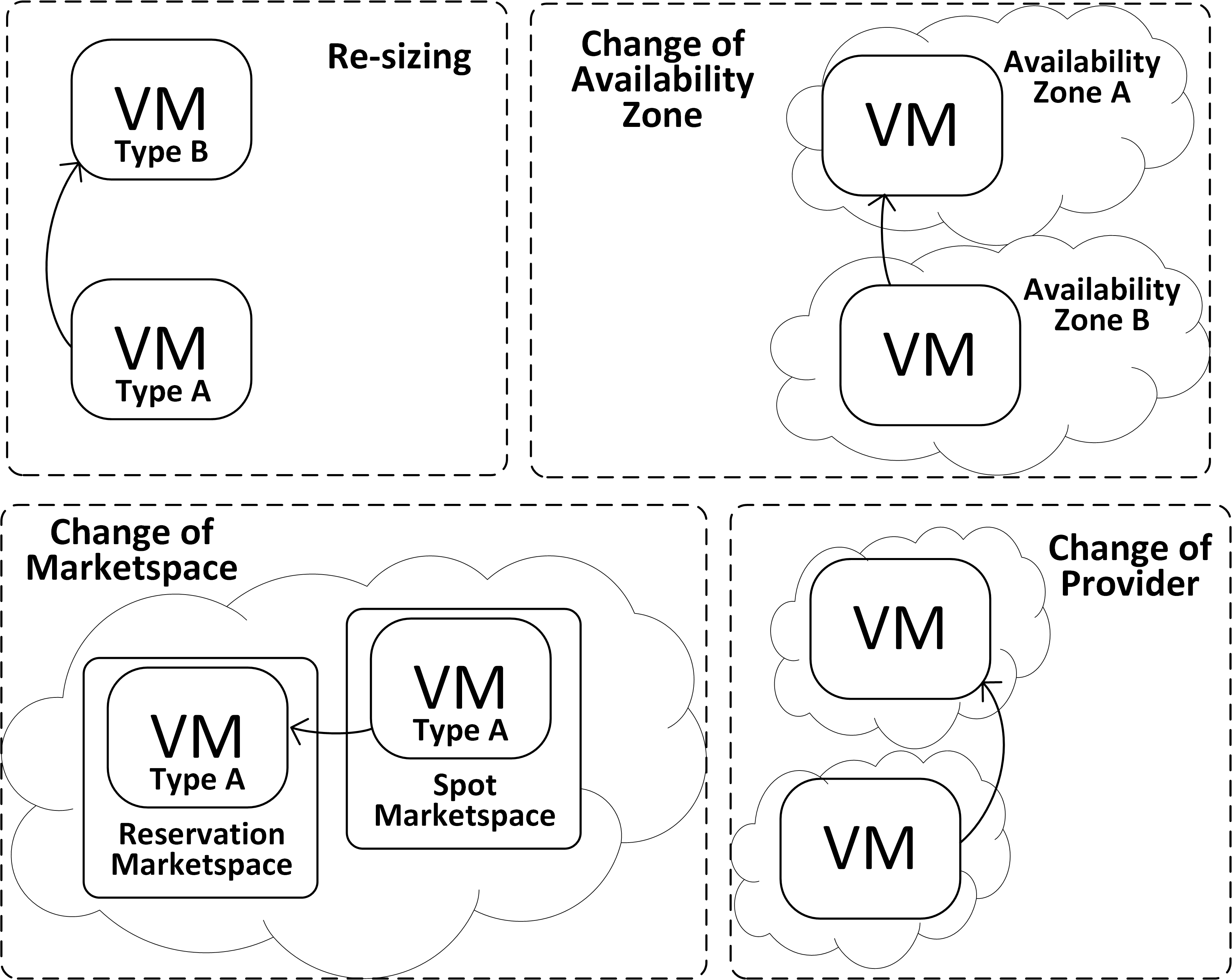}
	\subcaption{Different virtual machine migration scenarios in the cloud}
	\label{fig:migrationScenarios}
\end{subfigure}
\begin{subfigure}[c]{0.5\textwidth}
	\includegraphics[width=1\linewidth]{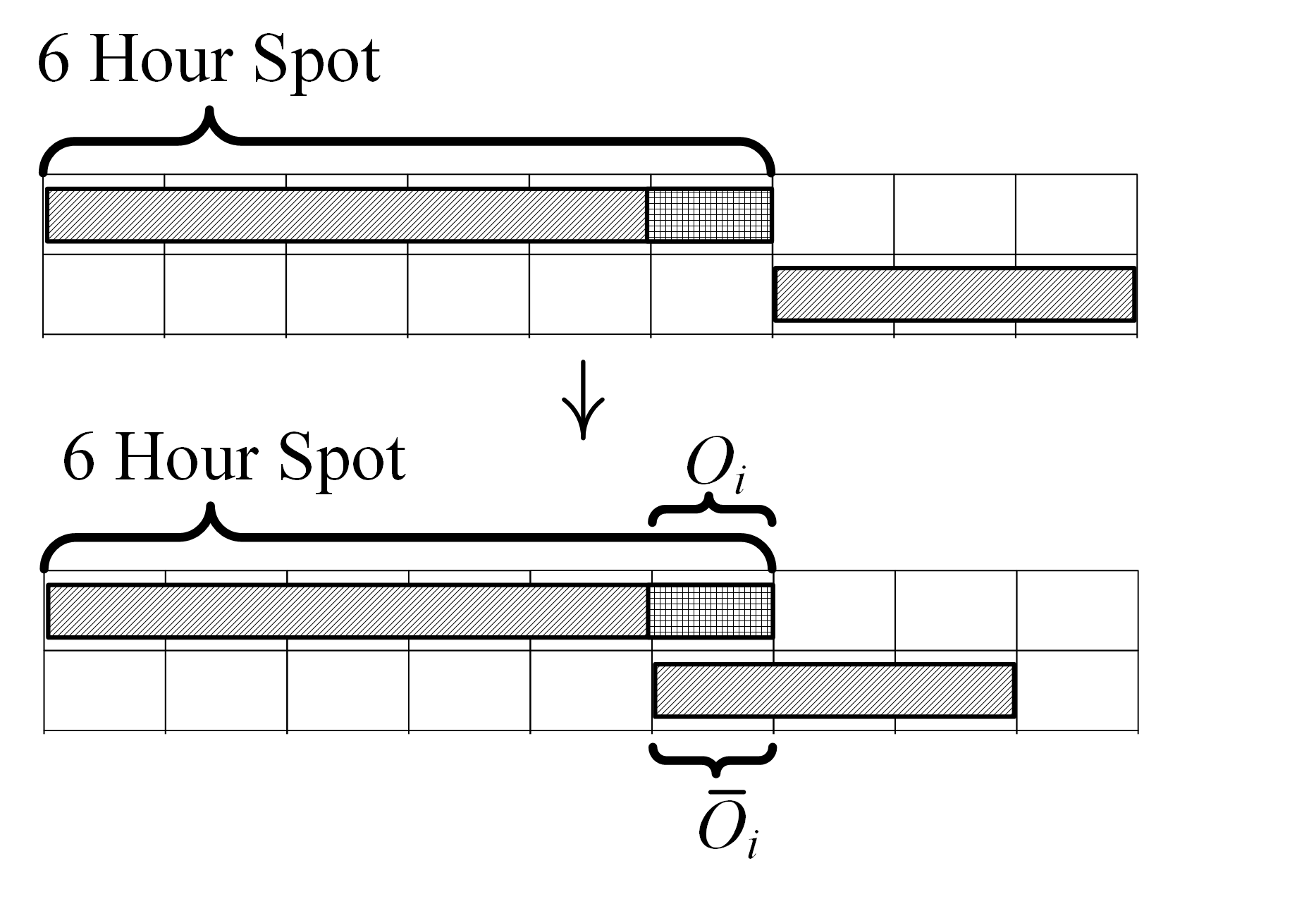}
	\caption{Migration between marketspaces requires time so that the virtual machines run at hour $i$ on both marketspaces}
	\label{fig:migrationScenario}
\end{subfigure}
\caption{Migrations of virtual machines}
\end{figure}

The term~\emph{Migration} is used for multiple movements of virtual machines in the cloud. The most important migrations are summarized in figure~\ref{fig:migrationScenarios}.
\begin{inparaenum}[(i)]
	\item Typically, a virtual machine can be migrated from one instance type to another instance type to fit the requested demand. Switching the instance types is called~\emph{re-sizing} on EC2. 
\item Migrating a virtual machine to another availability zone or marketspace requires the re-creation of the virtual machine in the new availability zone or on the new marketspace. The persistent data of the old virtual machine has to be transferred to the newly created virtual machine. This persistent data is usually stored on virtual hard disc drives - called devices on Amazon. Amazon distinguishes between EBS-backed devices and instance store-backed devices. 
The main difference between these two types is that instance store-backed devices are deleted after the virtual machine is terminated, while devices managed by EBS remain stored\footnote{https://docs.aws.amazon.com/AWSEC2/latest/UserGuide/ComponentsAMIs.html}. Two strategies for transferring the data between virtual machines are described in the following: 
	 \emph{EBS Image}.  The data of the old EBS device can be transferred to the newly created virtual machine by creating an EBS snapshot. This snapshot can be attached to the newly created virtual machine.
	 \emph{AMI}. An alternative to EBS images is the creation of so-called Amazon Machine Images (AMI). An AMI  contains all the information for running a virtual machine and includes, inter alia, the operating system. A migration between marketspaces requires time as a new virtual machine has to be started and the data from the old virtual machine has to be transferred to it. 
\item Figure~\ref{fig:migrationScenarios} further depicts the migrations of virtual machines between cloud providers. Such migrations are part of our further research but out of the scope of this paper, which focuses on migrations of virtual machines between marketspaces.
\end{inparaenum}

\section{Problem Formulation}
\label{sec:problem}

 The main challenge towards creating a cost-efficient cloud-portfolio is to find for a given set of requested virtual machines $x_{1}..x_{m}$ a set of virtual machines $v_{1}..v_{m}$ which are cost-efficient for a defined time interval $n$ and fulfill the requests as shown in equation~\ref{equ:1}.
\begin{equation}
	\label{equ:1}
	 x_{i} \implies \exists v_{i}: resource_{j}(x_{i}) \leq resource_{j}(v_{i}), \forall j, i \in 1..m
\end{equation}

$resource_{j}(x_{i})$ is a function that returns a resource $j$, such as memory or processing capacity, from a virtual machine or a request for a virtual machine.
On Amazon, a virtual machine $v$ is an instance of an instance type $\mathbf{T}$: $\bm{\mu}({v,\mathbf{T}})$. An instance type represents a pre-configured virtual machine that determines the available resources and technical characteristics. Using Amazon's instance types, $\mathbf{T}$ could take the following values: $\mathbf{T} \in \bm{\{}\text{t3.nano},\text{t3.micro},...\bm{\}}$. Exemplary instance types are given in table~\ref{tab:instances}. Each virtual machine $v$ can be purchased on different marketspaces $\mathbf{M}= \boldsymbol{\{} \mathcal{SM}, \mathcal{ODM}, 1\mathcal{HSM},6\mathcal{HSM}, 1\mathcal{YRM},$ $3\mathcal{YRM} \boldsymbol{\}}$ which determines a virtual machine's price. The symbols of $\mathbf{M}$, as well as the most important characteristics, are defined in table~\ref{tab:marketSymbols}. 
 So a virtual machine $v$ is characterized by its instance type $\mathbf{T}$ and its marketspace $\mathbf{M}$.  During the runtime, a virtual machine $v$ can be hosted on different marketspaces.  The first challenge is to determine the appropriate instance type for a requested virtual machine. The second challenge is to run the virtual machine on one or - with~\emph{migrations} - on multiple marketspaces so that a cost-efficient cloud-portfolio is formed. 

\begin{table}
\scriptsize
\caption{Marketspace Symbols}
\begin{center}
\begin{tabular}{p{1.2cm}p{5.5cm}p{1.8cm} p{1.8cm}}
\hline
\hline
\textbf{Symbol} &  \textbf{Name}  &  \textbf{Duration}  &  \textbf{Pricing}  \\
\hline
$\mathcal{SM}$ &  Spot Marketspace (can be interrupted) &  $\infty$ &  per Hour\\
$\mathcal{ODM}$	& On-Demand Marketspace & $\infty$ &  per Hour  \\
1$\mathcal{HSM}$	& 1-hour spot Marketspace & 1h &  per Hour  \\
6$\mathcal{HSM}$	& 6-hour spot Marketspace  & 6h &  per Hour \\
1$\mathcal{YRM}$	& 1-year reservation Marketspace  & 1 Year &  Fee  \\
3$\mathcal{YRM}$	& 3-year reservation Marketspace  & 3 Years &  Fee\\
\hline
\end{tabular}
\end{center}
\label{tab:marketSymbols}
\end{table}

The requests for virtual machines $x_{i \dots m}$, which are considered in that paper, are taken from the Bitbrains datacenter.
 We term the duration for which a portfolio and consequently its virtual machines are required~\emph{planning period}. On Amazon, consumers pay for any partial hour used\footnote{https://docs.aws.amazon.com/sdk-for-java/v1/developer-guide/tutorial-spot-instances-java.html} so that hours are an appropriate unit of measure for the planning periods.
 The goal is to find for each virtual machine $v_{j}$ of a portfolio the optimal marketspaces. First, the consumer has to decide from which marketspace a virtual machine $v_{j}$ should be purchased in the first hour of its planning period. This decision is termed $O_{1}^{j}$. For example, if it purchases the virtual machine from the 1-year reservation marketspaces it pays a fixed fee in return for 365-day access to the virtual machine. Alternatively, the consumer could purchase a virtual machine from the on-demand marketspace, which guarantees access to the virtual machine for one hour. After the consumer selected an appropriate marketspace for the first hour $O_{1}^{j}$, it has to redo the decisions for the following hours $O_{2...n}^{j}$ where $n$ is the planning period. For each virtual machine of a portfolio, $n^6$ possible solutions exist as $6$ marketspaces are considered. 
 $O^{*}_{j}$ is the vector of optimal decisions $O^{*}_{j}=(O_{1}^{j},O_{2}^{j},..O_{n}^{j})$ for the virtual machine $v_{j}$. 
 The function $m$ returns the selected marketspace of a decision $O_{i}^{j}$ at hour $i$: $m(O_{i}^{j}) \in  \boldsymbol{M}$. $c$ is a cost function that returns the costs of a certain decision at hour $i$: $c(O_{i}^{j})$.

 The optimal portfolios which are introduced in the following sections are portfolios with virtual machines for which  $O^{*}_{j}$ was determined so that the costs for the planning period of $n$ hours are minimized, whereby $c(O_{i})$ are the costs for running a virtual machine while $c(\bar{O}_{i})$ represent migration costs.
\begin{equation}
   \sum_{i=1}^{n} c(O_{i})+c(\bar{O}_{i}) \rightarrow min
\end{equation}

Homogeneous portfolios were created where all virtual machines are purchased from a single marketspace: $m(O_{o}^r)=m(O_{p}^s) \text{ }  \forall o,p \in  \bm{\{} 1 \dots n \bm{\}} \text{ }$ $\text{ } r,s \in \bm{\{} 1 \dots m \bm{\}}$. Then heterogeneous portfolios were created where virtual machines can be purchased from different marketspaces without considering migrations: $m(O_{o}^r)=m(O_{p}^r) \text{ }  \forall o,p \in  \bm{\{} 1 \dots n \bm{\}}, \text{ } r \in \bm{\{} 1 \dots m \bm{\}}$. Finally, heterogeneous portfolios were created where migrations are foreseen.
There are two main migration scenarios for virtual machines.

\begin{table}
\scriptsize
\caption{Amazon instance types available in Frankfurt}
\begin{center}
\begin{tabular}{p{1.7cm}p{1.7cm}p{1.7cm}p{1.7cm}p{1.7cm}p{1.7cm}p{1.7cm}p{1.7cm}p{1.7cm}p{1.7cm}p{1.7cm}}
\hline
\hline
\textbf{Name} &  \textbf{vCPU} & \textbf{ECU}  & \textbf{Memory (GiB)} & \textbf{Instance Storage}\\
\hline
t3.nano &  2 &  Variable  & 0.5GiB & EBS Only \\
t3.micro &  2 &  Variable  & 1GiB & EBS Only \\
...\\
\\
\hline
\end{tabular}
\end{center}
\label{tab:instances}
\end{table}

\begin{itemize}
	\item \textbf{Marketspace Switch.}
	Figure~\ref{fig:migrationScenario} depicts a scenario where a virtual machine is migrated to another marketspace $m(O_{i}) \neq m(O_{i+1})$ e.g. from the 6-hour spot marketspace to any other marketspace. During the execution of migration tasks, the virtual machine has to be hosted on both marketspaces to guarantee service continuation. As Amazon charges for partly used hours, we assume a migration time of one hour during which migration tasks are executed.  So, in case of a virtual machine from the  6-hour spot marketspace, the user can effectively use only 5 hours - 1 hour is used for migration tasks, and therefore redundant. So at the hour $i$ the virtual machine runs on the old marketspace $m(O_{i})$ and additionally on the new marketspace as the lower plot of figure~\ref{fig:migrationScenario} shows. We denote the decision for the new marketspace at the hour $i$ as $\bar{O}_{i}$. Such a decision $\bar{O}_{i}$ at hour $i$ exists if migration was executed. So $m(O_{i}) \neq m(O_{i+1}) \implies \exists \bar{O}_{i} \wedge m(\bar{O}_{i})=m(O_{i+1})$. 
	
	
	\item \textbf{Marketspace Extension.}
	Migrations are also necessary on marketspaces which offer virtual machines with a predefined lifetime: Virtual machines from the $1$-hour spot marketspace are inappropriate for migrations as they terminate after $1$ hour. Hence, this migration scenario is only relevant for the $6$-hour spot marketspace $6\mathcal{HSM}$.  If a consumer wants to use a virtual machine from the 6-hour marketspace for more hours, then it has to start a new virtual machine on the 6-hour marketspace. Similar to the marketspace switch, the consumer has to ensure that the lifespans of the old and the new virtual machines overlap to enable migration. 

\end{itemize}

The planning period can be structured into intervals $z^{a \rightarrow b}$. Such an interval represents the lifespan of a virtual machine on a marketspace and contains a set of decisions for the timeslots $a$ to $b$:
 	\begin{equation}
	  z^{a \rightarrow b} =
    \begin{cases}
      \boldsymbol{\{} \bar{O}_{a} \boldsymbol{\}} \cup \boldsymbol{\{}  O_{j} | a+1 \leq j \leq b \boldsymbol{\}}  & a>1  \\
       \boldsymbol{\{}  O_{j} | a \leq j \leq b \boldsymbol{\}}      & a=1
    \end{cases}
	\end{equation}
	

Within an interval, all the decisions have the identical marketspace so that the following condition is fulfilled: $\forall O_{i}, O_{j} \in z^{a \rightarrow b}: m(O_{i}) = m(O_{j})$. The function $m$ returns the marketspace of these decisions $m(z^{a \rightarrow b})=m(O_{i})$, whereby $O_{i} \in z^{a \rightarrow b}$. The costs of an interval represent the sum of the costs of decisions contained in the interval.

 	\begin{equation}
	   c(z^{a \rightarrow b}) =
    \begin{cases}
      c(\bar{O}_{a}) + \sum_{i=a+1}^{b} c(O_{i})       & a>1  \\
       \sum_{i=a}^{b} c(O_{i})       & a=1
    \end{cases}
	\end{equation}

 The number of decisions which an interval contains is limited by the contract period of the corresponding marketspace: $|z^{a \rightarrow b}| \leq t(m(z^{a \rightarrow b}))$.  
 At hour $1$, three different intervals are feasible: $z^{1 \rightarrow e}$, whereby $e$ represents any number $\le 8$, if the virtual machine is purchased from the $\mathcal{ODM}$ or the $\mathcal{SM}$ , $z^{1 \rightarrow 6}$ if the virtual machine is purchased from the $6\mathcal{HSM}$ and $z^{1 \rightarrow 8}$ if the virtual machine is purchased from the 1$\mathcal{YRM}$ or 3$\mathcal{YRM}$ marketspace. The 1$\mathcal{HSM}$ is not relevant for a planning period $>1$ hour as virtual machines on that marketspace terminate after 1 hour, which makes migrations impossible.
The previously described goal function can be rewritten as follows.

\begin{equation}
 \begin{split}
 \frac{c(z^{1 \rightarrow d})}{n}+ \frac{c(z^{d \rightarrow l})}{n}+ \dots + \frac{c(z^{l+j \rightarrow n})}{n} =  \\
	 \frac{c(z^{1 \rightarrow d})}{r^{1 \rightarrow d}} \cdot \frac{r^{1 \rightarrow d}}{n}+ \frac{c(z^{d \rightarrow l})}{r^{d \rightarrow l}} \cdot \frac{r^{d \rightarrow l}}{n}+ \dots +
	  \frac{c(z^{l+j \rightarrow n})}{r^{l+j \rightarrow n}}	\cdot \frac{r^{l+j \rightarrow n}}{n}=		 \\	
	  costs_{avg}^{1 \rightarrow d} \cdot \frac{r^{1 \rightarrow d}}{n}+ costs_{avg}^{d \rightarrow l} \cdot \frac{r^{d \rightarrow l}}{n}+ \dots +
	  costs_{avg}^{l+j \rightarrow n}	\cdot \frac{r^{l+j \rightarrow n}}{n}		 \\	 		
 \rightarrow min
 \end{split}
 \end{equation}

$r^{a \rightarrow b}$ represents the number of hours for which a virtual machine runs before a migration is executed. So the costs of an interval $c(z^{a \rightarrow b})$ are distributed over the non-overlapping timeslots $r^{a \rightarrow b}$ to calculate the average costs.
For example, the costs for the interval $z^{1 \rightarrow d}$ are distributed over $d-1$ timeslots:  $\frac{c(z^{1 \rightarrow d})}{d-1}$.  At the timeslot $d$ the virtual machine is running on two marketspaces for executing the migration. The costs of this overlapping timeslot are considered migration costs.
 Intervals have to be identified that lead to a minimum of average costs and so to an optimal solution.  
\begin{equation}
z^{*a \rightarrow b}: \nexists z^{a \rightarrow b+d} |
 costs_{avg}^{a \rightarrow b}  >  costs_{avg}^{a \rightarrow b+d}	
\end{equation}
$d$ represents any number; only the condition has to be fulfilled, which implies that an interval can not exceed the planning period: $b+d < n$.

\begin{figure}[ht!]
	\centering
	\includegraphics[width=0.50\linewidth]{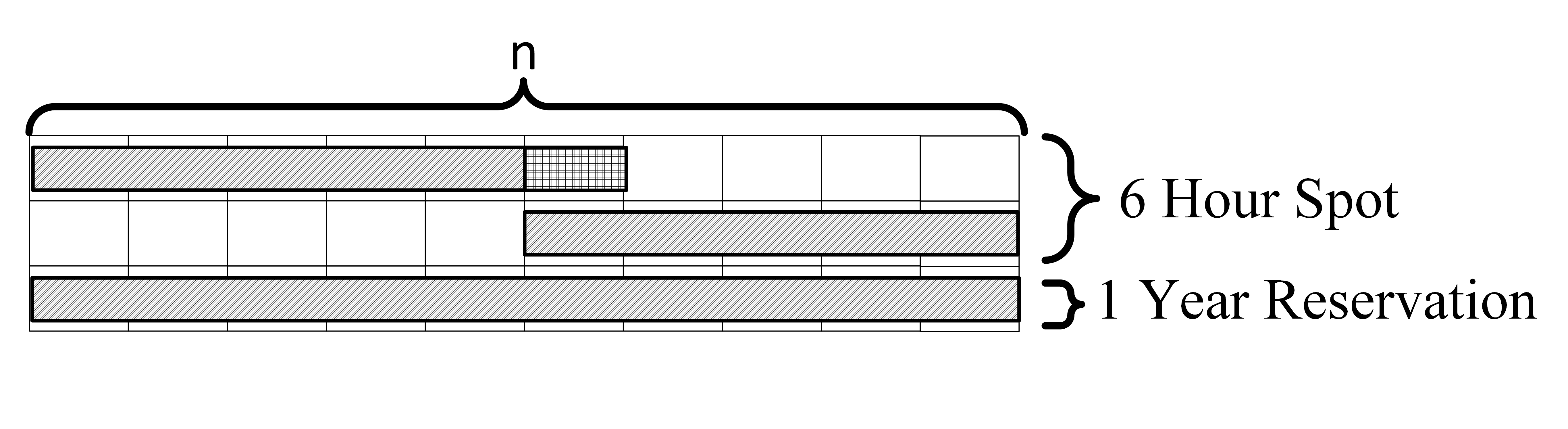}
	\caption{A scenario where the migration costs of a virtual machine from the 6-hour spot marketspace are distributed over $10$ timeslots}
	\label{fig:lastMile}
\end{figure}

A greedy-based algorithm can be used for finding an optimal solution. This algorithm uses the average costs for choosing the appropriate marketspace. This allows comparing the prices of the marketspaces $\mathcal{SM}$, $\mathcal{ODM}$ with the prices from the reservation marketspaces 1$\mathcal{YRM}$ and 3$\mathcal{YRM}$.  For the choice $O_{i}$ at the time $i$, the algorithm evaluates the marketspaces using the average costs  $costs_{avg}^{i \rightarrow i+l}$.  The average cost functions are summarized in table~\ref{tab:avgCosts}. There, it can be seen that the calculation of the average costs is based on the contract period and the remaining timeslots as the definition of $l$  shows: $1 \le l \le min(\text{remaining } \text{slots},t(m(O_{i}))-1) $ whereby the $remaining$ $slots$ are the number of timeslots for which a decision is pending:  $remaining$ $slots=n-i+1$.  $k$ is a variable that indicates whether a virtual machine from a marketspace can be used for the complete planning period or if at least one migration is necessary.   E.g., if a planning period is 370 days but the contract period of the reservation market is 365 days, then at least one migration from the 1-year reservation marketspace to any other marketspace is necessary so that $k=1$. If the planning period is 200 days, then the contract period of the 1-year reservation market is sufficient so that $k=0$. If a consumer uses the on-demand marketspace or the spot marketspace for the complete planning period, then no migration is necessary, and so $k=0$. A special situation has to be considered for the 6$\mathcal{HSM}$. At this marketspace consumers pay per hour for a virtual machine, but the lifetime is limited to 6 hours. If a migration is necessary, then the costs of the last timeslot, during which the migration is executed, are distributed over the remaining 5 timeslots. This requires special treatment of situations, which is, for example, simplified shown in figure~\ref{fig:lastMile}. There, the planning period $n$ is $10$. In the given example we assume that the average costs from the 6-hour spot marketspace $costs_{avg}^{{1} \rightarrow {6}}, k=1$ are higher than the average costs from the 1-year reservation marketspace $costs_{avg}^{{1} \rightarrow {10}}, k=0$:
\begin{equation}
	\begin{split}
	costs_{avg}^{{1} \rightarrow {6}}: c(O_{1})+\frac{c(O_{1})}{5},m(O_{1})=6\mathcal{YSM} > \\
	costs_{avg}^{{1} \rightarrow {10}}: \frac{c(O_{1})}{10},m(O_{1})=1\mathcal{YRM} > \\
	costs_{avg}^{{1} \rightarrow {6}, {6} \rightarrow {10}}: 0.5 \cdot \left(c(O_{1})+\frac{c(O_{1})}{5}\right)+  0.5 \cdot c(\bar{O}_{6}), \\ m(O_{1})=m(O_{6})=6\mathcal{YSM}
	\end{split}
\end{equation}	
	 If the average costs are used for choosing a marketspace then the 1-year reservation marketspace would be used for the $10$ timeslots as this interval has the lower average costs: $costs_{avg}^{{1} \rightarrow {6}}, m(O_{1})=6\mathcal{YSM} > costs_{avg}^{{1} \rightarrow {10}}, m(O_{1})=1\mathcal{YRM}$. However, considering the complete planning period of $10$ hours reveals that the 1-year reservation marketspace is not the optimal solution.  This is because no further migration is necessary for the second interval $costs_{avg}^{ {6} \rightarrow {10}}$ if the $6\mathcal{HSM}$ is used, and so the average costs decrease. This example shows that special cases exist where the average costs of using the 6-hour spot marketspace for the complete planning period $n$ are lower than the average costs of the 1-year reservation marketspace. Such a scenario is only possible if the average costs of the reservation marketspace are lower than the average costs from $6\mathcal{HSM}$ including a timeslot for migration, but higher than the average costs from $6\mathcal{HSM}$ excluding migration costs. To find the optimal solution, the algorithm has to compare the average costs of the $6\mathcal{HSM}$ with the average costs from $1\mathcal{YRM}$ and the $3\mathcal{YRM}$ by using the same interval length. So the algorithm calculates the costs of a virtual machine from the  $6\mathcal{HSM}$ for the same interval length, which is used for calculating the average costs on the reservation marketspace.

The steps of the algorithms are summarized in the following:
\begin{itemize}
	\item At timeslot $i$, the algorithm calculates the average costs for all marketspaces.
	\item The marketspace which offers the lowest average costs is used as long as possible:
 	\begin{equation}
    \begin{split}
      m(\bar{O}_{i})=\dots=m(O_{i+l})       \quad \text{if}  \quad a>1  \\
      m(O_{i})=\dots=m(O_{i+l})       \quad  \text{if} \quad a=1
    \end{split}
	\end{equation}	
	\item  After a marketspace was selected, the algorithm checks if all decisions for the planning period are defined. If $i+l<n$, then the algorithm re-evaluates for the remaining timeslots the available marketspaces using the average costs as described in the first step.

\end{itemize}

\begin{table}
\scriptsize
\caption{Average Costs Functions}
\begin{center}
\begin{tabular}{p{3.5cm}p{7cm}}
\hline
\hline
\textbf{Marketspace} &  \textbf{Average Costs Functions $costs_{avg}^{i \rightarrow i+l}$} ($O^{*}$)  \\
\hline
$\mathcal{SM}$      &  $c(\hat{O}_{i})$ + $ \frac{c(\hat{O}_{i})}{l}  \cdot k $\\
$\mathcal{ODM}$	    &  $c(\hat{O}_{i})$ + $ \frac{c(\hat{O}_{i})}{l} \cdot k $ \\
1$\mathcal{HSM}$	& not relevant \\
6$\mathcal{HSM}$	& $c(\hat{O}_{i}) + \frac{c(\hat{O}_{i})}{5} \cdot k $ \\
1$\mathcal{YRM}$	& $\frac{c(\hat{O}_{i})}{l +(1-k)}$  \\
3$\mathcal{YRM}$	& $\frac{c(\hat{O}_{i})}{l + (1-k)}$  \\
	& $\hat{O}_{i}$=$\bar{O}_{i}$ if $\exists$ $\bar{O}_{i}$\\
	& $\hat{O}_{i}$=$O_{i}$ if $\nexists$ $\bar{O}_{i}$\\
\hline
\end{tabular}
\end{center}
\label{tab:avgCosts}
\end{table}

%
%
%
%
%
%
%
%
%

 Algorithm~\ref{alg:costOpti} details the previously described steps.
\begin{algorithm}
  \caption{Marketspace Optimization - simplified}\label{alg:costOpti}
  \begin{algorithmic}[1]
    \Procedure{Optimize}{$n$}\Comment{planning period n}
      \State i$\gets$0 \Comment{current timeslot}
      \State $O^{tmp}$ \Comment{temporary timeslot, $\notin O^{*}$}
      \While{$i<n$}
        \State k$\gets$0
        \State $occupiedSlots$ $\gets$0
        \State $localMinimum \gets \infty$
        \State $remainingSlots \gets n$-$i$+$1$
	    \For{m $\in$ $\boldsymbol{M}$}
			\State $m(O^{tmp}) \gets m$ \Comment{set marketspace to m}
			\If {$remainingSlots$-$t(m(O^{tmp}))>$ 0 $\wedge$
			\\\hspace{1.8cm} $m \in$ $\boldsymbol{\{}$ 1$\mathcal{HSM}$,6$\mathcal{HSM}$,1$\mathcal{YRM}$,3$\mathcal{YRM}$ $\boldsymbol{\}}$  			
			\\\hspace{1.5cm}}
          		\State k $\gets$ 1 \Comment{migration is necessary}
        	\EndIf
        	\If {$m(O^{tmp})$ $\in$ $\boldsymbol{\{}$ $\mathcal{SM}$,$\mathcal{ODM}$ $\boldsymbol{\}}$ \\\hspace{1.5cm}} \\\hspace{1.8cm}
				$l$=0
			\Else  \\\hspace{1.8cm}
				$l$=$min(remainingSlots,t(m(O^{tmp})))-1$	
			\EndIf
			\State avgCosts $\gets$ $costs_{avg}^{i \rightarrow i+l}$ ($O^{*} \cup O^{tmp}_{i}$) \\ \Comment{get avg costs using the equation from table~\ref{tab:avgCosts}}
			\If {$avgCosts<$ $localMinimum$}
          		\State $localMinimum \gets avgCosts$
          		\State $localMinimumMarket \gets m$
          		\State $occupiedSlots$ $\gets l$
        	\EndIf
        \EndFor		

        \If {$O_{i} \in O^{*}$}
        	\State $O^{*}$=$O^{*}$ $\cup$ $\bar{O}_{i}$
        \EndIf
        \For{j $\in$ 0 $:$ $occupiedSlots$ }
        	\State $m(\hat{O}_{i+j}) \gets localMinimumMarket$
        	\State $O^{*}$=$O^{*}$ $\cup$ $\hat{O}_{i+j}$
        \EndFor

        \State $i \gets i$+l+$1$-$k$
      \EndWhile
    \EndProcedure
  \end{algorithmic}
\end{algorithm}


\section{Dataset and Instance Type Mapping}
\label{sec:instancetypemapping}

The dataset from the Bitbrains datacenter contains the utilization of virtual machines and is structured into two sub-datasets:
\begin{itemize}
	\item \emph{Fast Storage}: Here, 1250 virtual machines were traced over one month. 
	\item \emph{RND}: In this dataset, 500 virtual machines were traced over three months. 
\end{itemize}

The two datasets represent groups of virtual machines which were used for different fields of applications: Virtual machines of the first dataset were used as application servers, while virtual machines of the second dataset were used as management machines. All virtual machines are business-critical. Further details, as well as an analysis of the workloads without considering pricing aspects, are presented in~\cite{ShenBI15}. The dataset consists of CSV files, where the utilization of each virtual machine is stored in a separate file. An excerpt from such a file is given in table~\ref{tab:csvfile}. It can be seen that the utilization of the disk, memory, and the CPU of each virtual machine is traced (columns CPU usage, memory usage, disk read throughput,...). In addition to the utilization, the file contains the resources of the virtual machine which were requested by the consumers (columns CPU capacity, Memory capacity,...).  
We assumed that all requested virtual machines should be hosted in central Europe, and so in Amazon's availability zone Frankfurt.

\begin{table*}
\scriptsize
\caption{Head of the trace file (fast storage, 2013-8-1)}
\begin{center}
\begin{tabular}{p{1.5cm}p{.8cm}p{0.9cm}p{0.9cm}p{0.9cm}p{1.cm}p{1.cm}p{1.cm}p{0.5cm}}
\hline
\hline
\textbf{Timestamp} &  \textbf{CPU cores} & \textbf{CPU capacity}  & \textbf{CPU usage [MHz]} & \textbf{CPU usage [\%]} & \textbf{Memory capacity} & \textbf{Memory usage} & \textbf{Disk read throughput [KB/s]} & \textbf{...} \\
\hline
1376314846 &  4 &  11704.00  & 10912.03 & 93.23 & 67108864 & 6129274.4 & 0.133 & ... \\
1376315146	& 4 &  11704.00 &	10890.57 & 93.05 &67108864 & 6755624.0 & 1.33 & ...  \\
...\\
\\
\hline
\end{tabular}
\end{center}
\label{tab:csvfile}
\end{table*}


The following sections focus on the~\emph{Fast Storage} dataset. The~\emph{RND} dataset is used for the validation of our results.
In a first step, the resources of the virtual machines from the Bitbrains dataset (columns CPU capacity, memory capacity of table~\ref{tab:csvfile}) were treated as requests for virtual machines. For those requests, appropriate instance types on Amazon were identified to create cloud portfolios. An appropriate instance type is an instance type that provides at least the same resources as the request, e.g., RAM - see equation~\ref{equ:1}. 
 Therefore, the instance type characteristics (see table~\ref{tab:instances}) and the corresponding CPU clock speed data\footnote{https://aws.amazon.com/ec2/instance-types/} were used. The disk storage is managed by Amazon independently of the instance types.  So it was not considered for finding appropriate instance types.
From all available instance types of the availability zone Frankfurt, we excluded the instance types that are not available on all marketspaces, such as the instance type $u-12tb1.metal$. Further, we excluded instance types such as~\emph{p2.8xlarge} for which Amazon did not publish the frequency of a virtual machine interruption on the spot marketspace, which is used in the following for calculating portfolio costs.  In total $80$ instance types with a Linux-based operating system were considered.

\begin{figure*}
	\centering
	\includegraphics[width=1\linewidth]{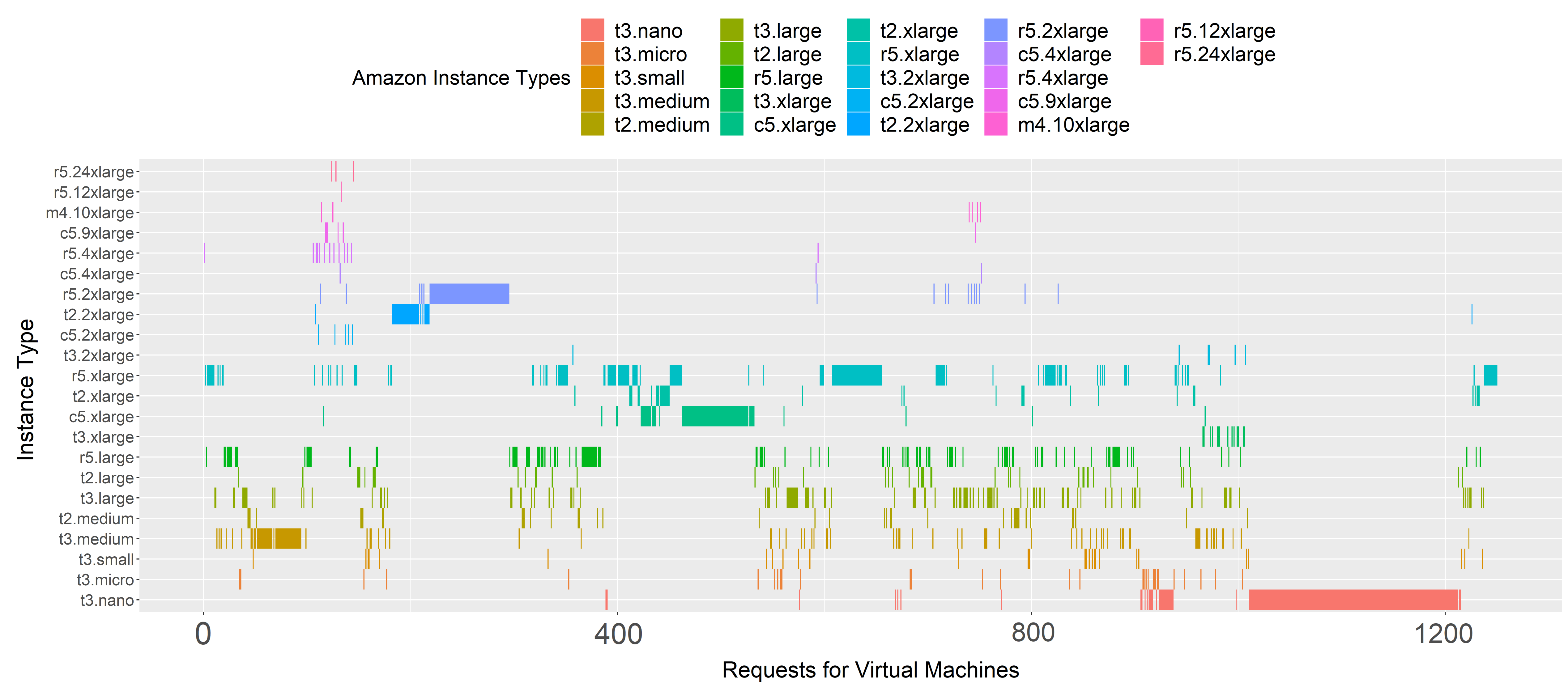}
	\caption{Cloud-portfolio resulting from the mapping of the requests for virtual machines to instance types available on Amazon \emph{using the requested resources}}
	\label{fig:instancevmmapping}
\end{figure*}

The results of the mapping process are shown in figure~\ref{fig:instancevmmapping}. The abscissa of the figure shows the requests from the Bitbrains dataset. For example, $1$ represents the first request for a virtual machine of the Bitbrains dataset, while $1250$ represents the last request for a virtual machine. The ordinate shows the corresponding instance type on Amazon, ordered by price. $t3.nano$ is the cheapest instance type, while the instance type $r5.24xlarge$ is the most expensive instance type. So for the first request, the instance type $r5.4xlarge$ is the cheapest matching instance type, for the requests $1100$ to $1200$, the instance type $t3.nano$ is the cheapest matching instance type. Table~\ref{tab:instancesUsed} shows the results in more detail (column requested). It can be seen that the requests are distributed over $20$ different instance types. For most of the requests, virtual machines of the type $t3.nano$ are sufficient.

\begin{figure*}
	\centering
	\includegraphics[width=0.99\linewidth]{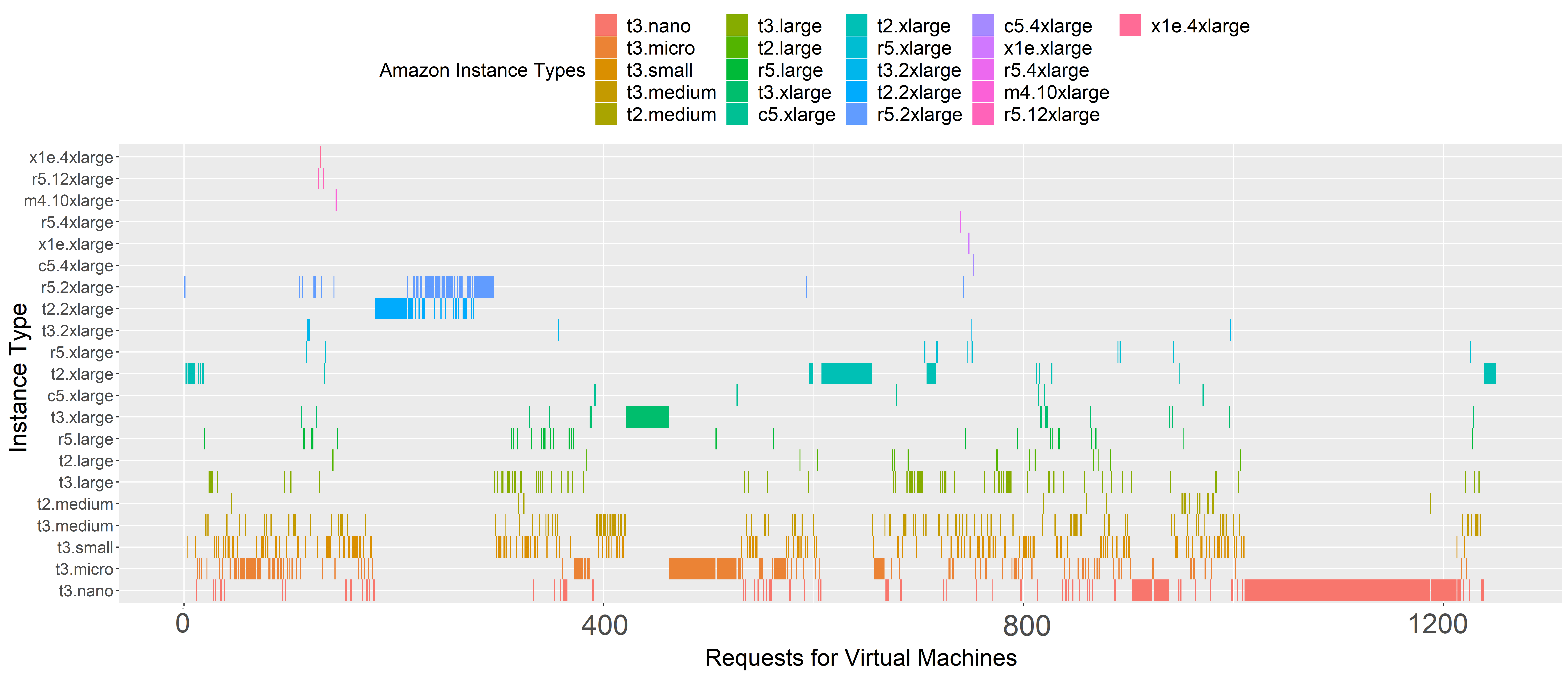}
	\caption{Cloud-portfolio resulting from the mapping of the requests for virtual machines to instance types available on Amazon \emph{using the maximum utilization of the virtual machines}}
	\label{fig:instancevmmappingutilization}
\end{figure*}

We created a second mapping whereby we used the maximum utilization of a virtual machine instead of the requested resources (maximum of columns CPU usage, memory usage of table~\ref{tab:csvfile}) for the mapping process. The result is visualized in figure~\ref{fig:instancevmmappingutilization}. It can be seen that the density of small instance types (bottom of the plot) is higher than the density of small instances in figure~\ref{fig:instancevmmapping}.  This is because most of the virtual machines are over-sized, which means that their resources are never used. Indeed, $888$ virtual machines (71\%) are underutilized and can be downgraded to smaller instance types. This represents a significant potential for cost savings. Table~\ref{tab:instancesUsed} shows the differences between the two mapping processes. For example, the expensive instance type $r5.24xlarge$ is not needed in the portfolio created with the maximum utilization, as its resources are never used, while the number of virtual machines based on the cheap instance type $t3.nano$ increases.

\begin{table}[ht!]
\scriptsize
\caption{Matching Amazon instance types for the requests}
\begin{center}
\begin{tabular}{p{3.1cm}p{2cm}p{3.0cm}p{1cm}}
\hline
\hline
\textbf{Types} &  \textbf{Requested} & \textbf{Maximum Utilization} & \textbf{$\Delta$}  \\
\hline
t3.nano & 235 & 314 & 79  \\
t3.micro & 31 & 207 & 176 \\
t3.small & 28 & 168 & 140  \\
... & ... & ... & ...  \\
r5.4xlarge & 13 & 1 & 12  \\
c5.9xlarge & 6 & 0 & 6  \\
m4.10xlarge & 6 & 1 & 5  \\
r5.12xlarge & 1 & 2 & 1  \\
x1e.4xlarge & 0 & 1 & 1  \\
r5.24xlarge & 3 & 0 & 3 \\
\hline
\hline
Total & 1250 & 1250 & 1038 \\
\hline
\end{tabular}
\end{center}
\label{tab:instancesUsed}
\end{table}

The portfolio which resulted from the second mapping process - using the maximum utilization - leads to a stronger concentration of instance types according to the Gini-index ($	Gini-index=\sum_{i=1}^{q}{(k_{i-1}+k_{i}) \frac{a_{i}H_{i}}{\sum_{j=1}^{q}{a_{j} H_{j}}}-1}$~\cite{gini1921measurement}):, i.e. few instance types are used for most of the requests. The first mapping process leads to a corrected Gini-index\footnote{see appendix for more information} of 0.61, while the second mapping process leads to a corrected Gini-index of 0.69. 


\section{Homogeneous Portfolios}
\label{sec:homoPort}

Virtual machines of homogeneous portfolios are purchased from a single marketspace. The hourly costs of such homogeneous portfolios are depicted in table~\ref{tab:portfolioCost}\footnote{prices taken from 08/2019}. Portfolios, where the instance types of the virtual machines were determined based on the maximum utilization instead of the requested resources, lead to significant cost savings on all marketspaces, e.g., for the 1-hour spot marketspace these cost savings are approximately 44\%.   For the reservation marketspaces (1 year and  3 years), consumers have to pay a fixed fee. The costs per hour which are given in table~\ref{tab:portfolioCost} are calculated by distributing the fee of the 1-year reservation marketspace and the 3-year reservation marketspace over the complete runtime: $\text{costs per hour}=\frac{fee}{365 \cdot 24}$ and $\text{costs per hour}=\frac{fee}{365 \cdot 24 \cdot 3}$. The on-demand marketspace has the highest costs, while the spot marketspace has the lowest costs.

Amazon offers low prices on the spot marketspace in return for the right to reclaim the resources at any time. The frequency of such an interruption is published on Amazon's website\footnote{https://aws.amazon.com/ec2/spot/instance-advisor/} for each instance type and availability zone. There, it is defined as follows: \emph{Frequency of interruption represents the rate at which Spot has reclaimed capacity during the trailing month. They are in ranges of \textless 5\%, 5-10\%, 10-15\% ,15-20\% and \textgreater 20\%.}
 The interruption rate published on Amazon can be interpreted as an arrival rate $\lambda$, which is widely used for Poisson processes. For example, for the instance type~\emph{t3.micro}, the monthly arrival rate of an interruption is $\lambda_{month} = 5\%$.   Based on the given monthly arrival rate, the hourly arrival rate $\lambda_{hour}$ can be determined. It represents the expected number of interruptions of a virtual machine of a certain instance type per hour. For consumers, an interruption is unpredictable and leads to unexpected service interruptions and consequently costs. So consumers of virtual machines from the spot marketspace have to pay the spot price to Amazon, and additionally, they have to consider costs for service interruptions or counter-measures preventing service interruptions. In the paper, these costs are represented by penalties. Hence, the expected costs of an interruption per hour are defined by multiplying the hourly arrival rate by the penalty, as the following equation shows.

\begin{table}
\scriptsize
\caption{Costs per hour of homogeneous portfolios without penalties}
\begin{center}
\begin{tabular}{p{5.cm}p{1.8cm}p{1.8cm}p{1.3cm}}
\hline
\hline
\textbf{Marketspace} &  \textbf{Requested} & \textbf{Maximum Utilization} & \textbf{$\Delta$}    \\
\hline
On-demand  Marketspace &  277.67\$  &  148.32\$  & 129.35\$  \\
Spot Marketspace  &  73.83\$ & 41.06\$ & 32.77\$    \\
1-hour Spot Marketspace  &  140.32\$  & 76.4\$  & 63.92\$   \\
6-hour Spot Marketspace  & 181.41\$  & 97.99\$  & 83.42\$   \\
No Prepaid - Reserved Marketspace (1 Year)  &  182.18\$ &   100.35\$ & 81.83\$  \\
Partly Prepaid - Reserved Marketspace (1 Year)  &  173.84\$ &  95.83\$ & 78.01\$    \\
Prepaid - Reserved Marketspace (1 Year)  &  170.18\$ &  93.81\$ & 76.37\$   \\
No Prepaid - Reserved Marketspace (3 Years)  &  130.85\$ &  71.75\$  & 59.1\$  \\
Partly Prepaid - Reserved Marketspace (3 Years)  &  121.89\$  &  66.56\$  & 55.33\$  \\
Prepaid - Reserved Marketspace (3 Years)  &  119.02\$ & 65.95\$ & 53.07\$    \\
\hline

\end{tabular}
\end{center}
\label{tab:portfolioCost}
\end{table}

\begin{figure*}[ht!]
	\centering
	\includegraphics[width=0.9\linewidth]{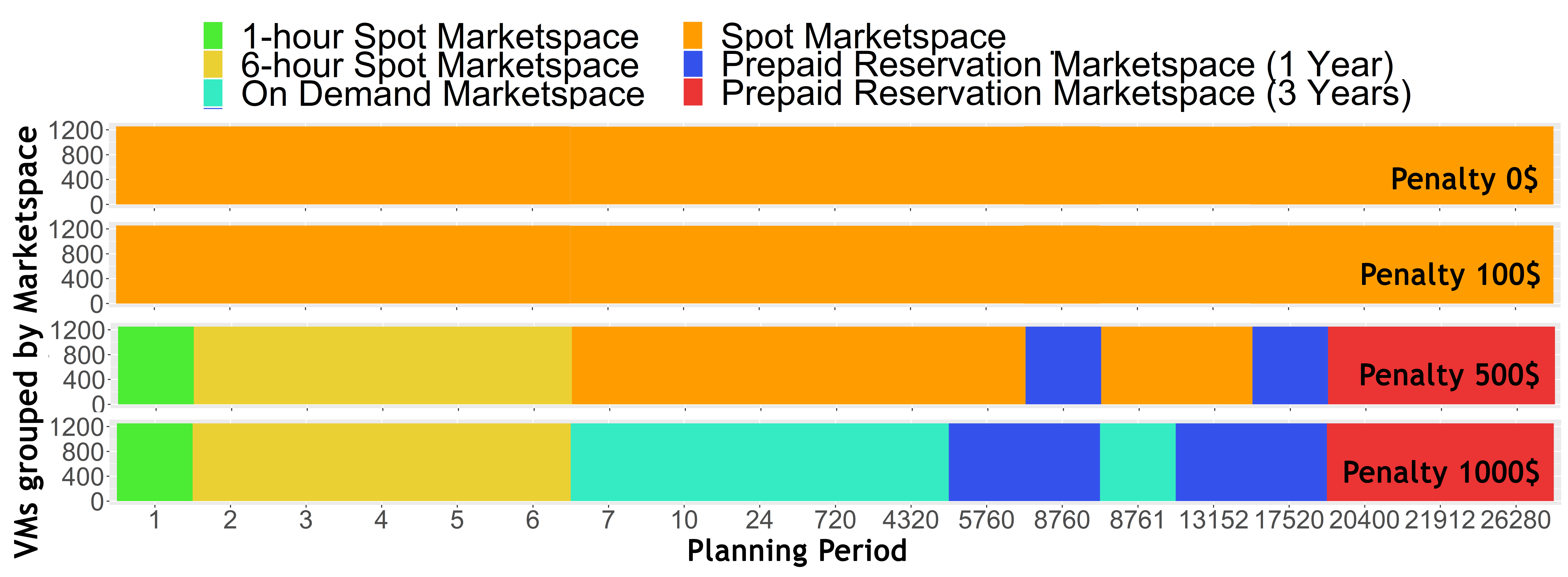}
	\caption{Optimal homogenous portfolios (all virtual machines are hosted during the planning period on the same marketspace) for different penalties: $0\$$, $100\$$, $500\$$, and $1000\$$}
	\label{fig:staticPortfolios}
\end{figure*}


\begin{equation}
	penalty_{hour}=\lambda_{hour} \cdot penalty
\end{equation}

So the costs per hour for a virtual machine on the spot marketspace contain the price paid to Amazon and the expected costs of an interruption as determined in the following equation.
\begin{equation}
	costs_{hour}=\text{spot price} + penalty_{hour}
\end{equation}


Figure~\ref{fig:staticPortfolios} shows optimal homogeneous portfolios for different penalties and planning periods.  All virtual machines of homogeneous portfolios are hosted on a marketspace during the complete planning period (no migrations).  The abscissa shows the planning periods in hours, while the ordinate shows the virtual machines of the portfolios grouped by marketspace. Each tick on the abscissa represents a portfolio with a certain planning period. 
The first plot of the figure shows portfolios without considering penalties. Without any penalty, portfolios where all virtual machines are purchased from the spot marketspace lead to the lowest costs for all planning periods. The same portfolio structure results when a penalty of $100\$$ is used as the second plot in figure~\ref{fig:staticPortfolios} shows.  
The third plot of figure~\ref{fig:staticPortfolios} shows portfolios where a penalty of $500\$$ is assumed. For a planning period of $1$ hour, all virtual machines of the optimal portfolio are purchased from the 1-hour spot marketspace. Virtual machines from the 1-hour spot marketspace terminate after $1$ hour and so they are not relevant for longer planning periods. For the planning periods $2$ hours to $6$ hours, an optimal portfolio consists of virtual machines that are purchased from the 6-hour spot marketspace. Virtual machines from the 6-hour spot marketspace terminate after $6$ hours, and so they are not relevant for longer planning periods. Hence, the spot marketspace becomes relevant for portfolios with a planning period from $7$ to $5760$ hours ($240$ days). For a planning period of $8760$ hours, the 1-year reservation marketspace is necessary to host the portfolio at the lowest costs. The spot marketspace becomes relevant again to form optimal portfolios with the planning periods $1$ year and $1$ hour, and $1.5$ years. This results from the inappropriate contract period of the 1-year reservation marketspace: for portfolios with a planning period of $8761$ hours and  $13152$ hours (1 year and 1 hour and $1.5$ years) the fee for the 1-year reservation marketspace would have to be paid two times as the contract period of a virtual machine of the 1-year reservation marketspace is  $1$ year.  For the remaining planning periods (2 years and 120 days), the reservation marketspaces are necessary to host the portfolio at the lowest costs. The last plot in figure~\ref{fig:staticPortfolios} shows the optimal portfolio structure when a penalty of $1000\$$ is assumed. The structure of the optimal portfolios is different than before at the planning periods $7$ to $4320$ hours, $8761$ hours and $13152$ hours: For these planning periods, the on-demand marketspace, as well as the 1-year reservation marketspace, is used instead of the spot marketspace due to the increased prices resulting from the penalty of $1000\$$.

%
%

\section{Heterogeneous Portfolios}
\label{sec:costAnalysis}

\begin{figure*}[ht!]
	\centering
	\includegraphics[width=0.95\linewidth]{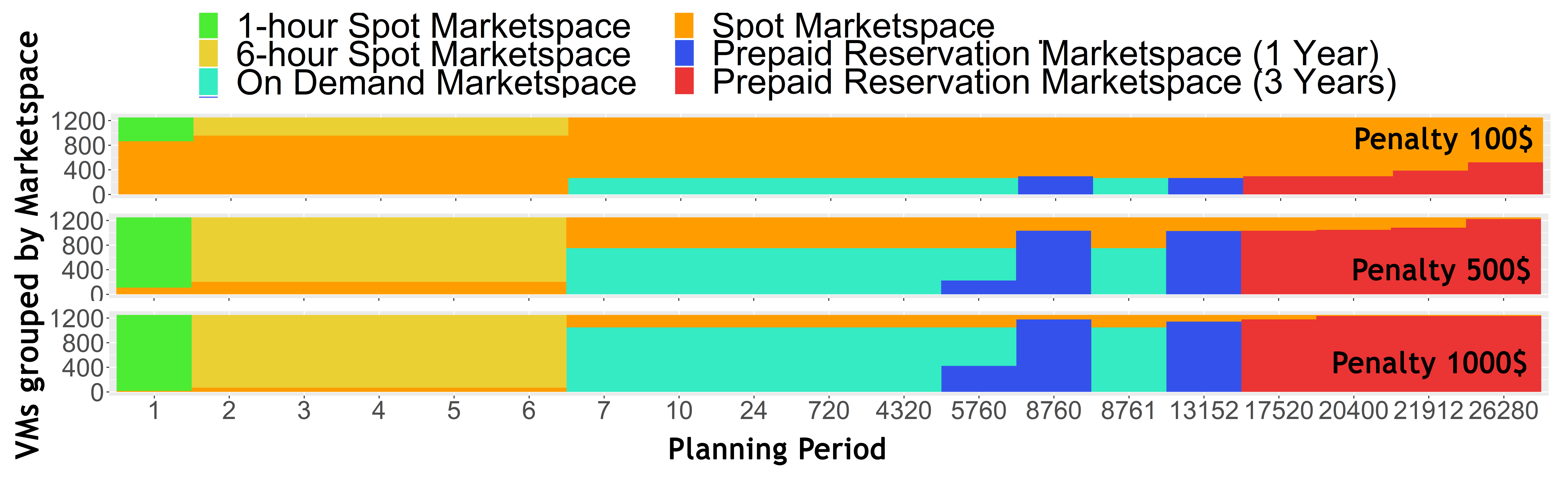}
	\caption{Optimal heterogenous  portfolios for different penalties: $100\$$, $500\$$ and $1000\$$, without considering migration}
	\label{fig:portofliosWithotuMigration}
\end{figure*}

Purchasing the virtual machines of a portfolio from a single marketspace does not necessarily lead to the lowest costs. The following portfolios foresee that virtual machines can be purchased from different marketspaces, leading to heterogeneous cloud portfolios. Such portfolios are depicted in figure~\ref{fig:portofliosWithotuMigration} where virtual machine migrations are neglected. The upper plot of the figure shows the optimal portfolios when a penalty of $100\$$ is assumed. For the shortest planning period, which is $1$ hour, most of the virtual machines are purchased from the 1-hour spot marketspace. The remaining virtual machines are purchased from the spot marketspace. For longer planning periods, the virtual machines are purchased from the 6-hour spot marketspace, the rest from the spot marketspace. After six hours, virtual machines from the 6-hour marketspace can not be used without migration. So for portfolios that have to be hosted between 7 and 4320 hours, virtual machines have to be purchased from the on-demand marketspace and the spot marketspace. For a planning period of 1 year, some of the virtual machines have to be purchased from the 1-year reservation marketspace to create optimal cloud portfolios. An interesting portfolio structure is formed for a planning period of 1 year and 1 hour: $8761$ hours. Similar to the corresponding homogeneous portfolio, all its virtual machines have to be purchased from the on-demand marketspace and the spot marketspace, but not from the 1-year reservation marketspace. This is because the contract period of virtual machines from the 1-year reservation marketspace is limited to $8760$ hours, and so consumers would have to pay the fee of the 1-year reservation marketspace twice.  For the remaining planning periods, optimal portfolios consist of virtual machines that are purchased from the spot marketspace and the 3-year reservation marketspace. Figure~\ref{fig:portofliosWithotuMigration} shows two further plots that represent the optimal portfolios when using a penalty of $500\$$ and $1000\$$ for an interruption of a virtual machine on the spot marketspace. The portfolio structure is almost identical for all three scenarios; only the share of the spot marketspace decreases with an increasing penalty. In table~\ref{tab:averageCosts}, the average costs per virtual machine per hour are listed. The heterogeneous portfolios without migration (columns heterogeneous portfolio) outperform the corresponding homogeneous portfolios for all planning periods.

\begin{figure*}[ht!]
	\centering
	\includegraphics[width=0.95\linewidth]{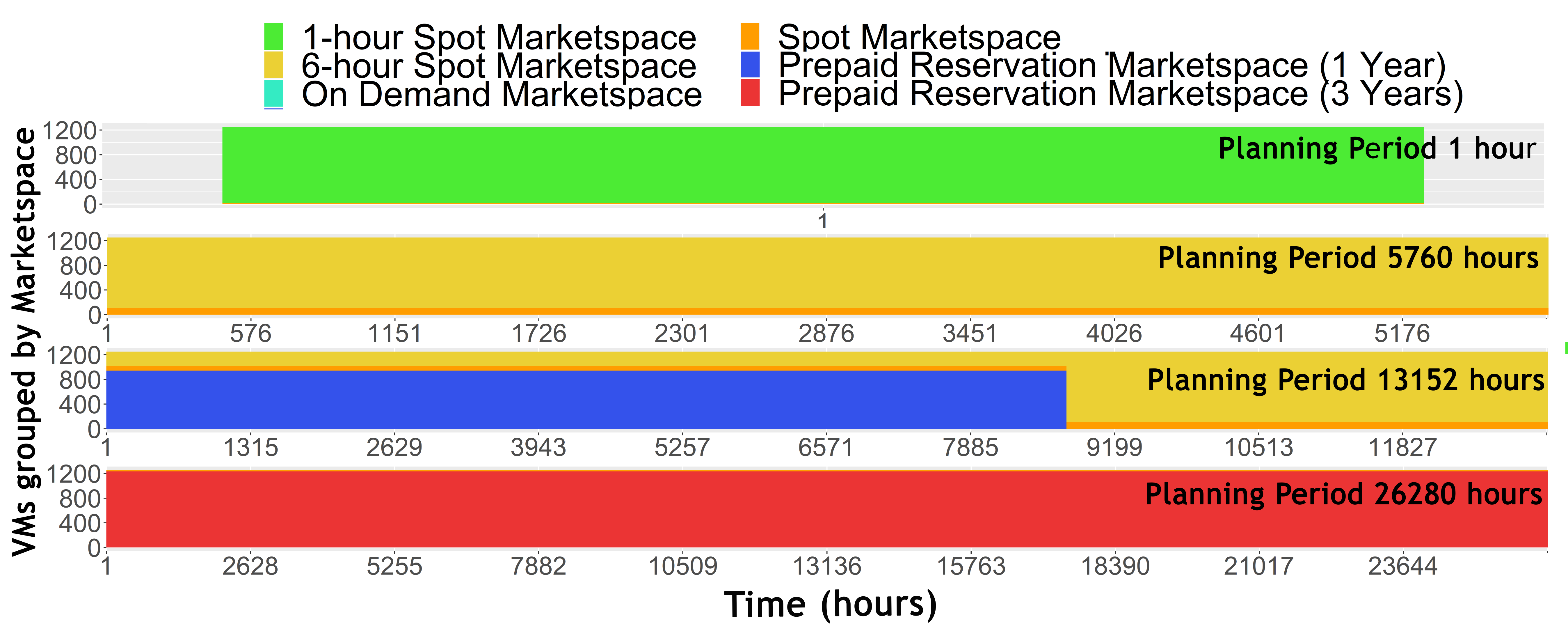}
	\caption{Portfolio structure for the planning periods $1$, $5760$, $13152$ and $26280$ hours considering a penalty of $1000\$$, with migration}
	\label{fig:portfolioMigration1000}
\end{figure*}

With the migration techniques described in section~\ref{sec:foundations}, consumers could move virtually to different marketspaces during runtime to further improve the portfolio structure. 
Optimal portfolios where migration is allowed are depicted in figure~\ref{fig:portfolioMigration1000}. In that figure, the abscissa does not represent the planning period. Instead, it represents the portfolio structure at a certain point in time. In the first plot of the figure, the planning period is $1$ hour and no migration is executed. So the structure of that portfolio is identical to the structure of the portfolio with a penalty of $1000\$$ at planning period $1$ in figure~\ref{fig:portofliosWithotuMigration}. In figure~\ref{fig:portfolioMigration1000}, the optimal portfolio with a planning period of $5760$ hours contains virtual machines from the spot marketspace and the 6-hour spot marketspace. The share of virtual machines from the 6-hour marketspace remains constant during the runtime.  This implies migrations that have to be done for all $6$ hours. For those virtual machines, it is more cost-efficient to run and migrate them on the 6-hour spot marketspace than running them on the spot marketspace or the on-demand marketspace, which was foreseen for the optimal portfolio without migration  (planning period of $5760$) - see figure~\ref{fig:portofliosWithotuMigration}. The third plot of figure~\ref{fig:portfolioMigration1000} shows the optimal portfolio for the planning period of $13152$ hours. It can be seen that migrations to other marketspaces are necessary. So for the first year ($8760$ hours), some of the virtual machines are hosted on the 1-year reservation marketspace. For the remaining $183$ days, these virtual machines are migrated to the 6-hour spot marketspace. 
The last plot in figure~\ref{fig:portfolioMigration1000} shows an optimal portfolio with a planning period of 3 years, where no migration is necessary.


\section{Portfolio Cost Analysis}
\label{sec:cost}

\begin{figure*}[ht!]
	\centering
	\includegraphics[width=0.99\linewidth]{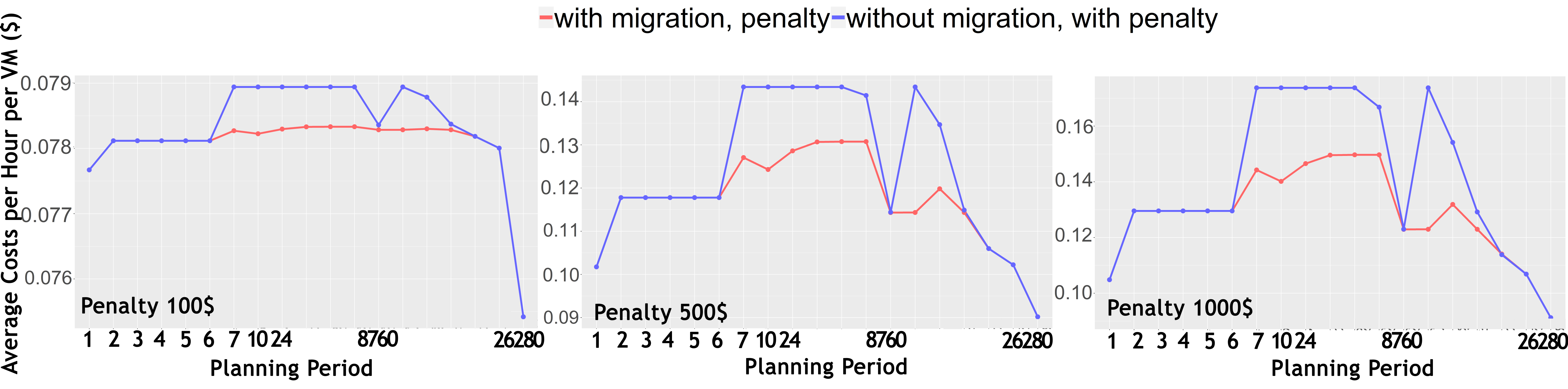}
	\caption{Average costs per hour per virtual machine of the portfolios with migration and without migration using $100\$$ penalty (left), $500\$$ penalty (center), and $1000\$$ penalty (right) - see appendix for the data }
	\label{fig:avgcosts}
\end{figure*}

Figure~\ref{fig:avgcosts} depicts the average costs per hour per virtual machine of optimal portfolios, which allows an analysis of the cost differences between the portfolios with migration (red line) and without migration (blue line). It is a visualization of the data from table~\ref{tab:averageCosts}. The average costs per hour per virtual machine are low for short planning periods, where virtual machines from the 1-hour and 6-hour spot marketspace can be used, as well as for long planning periods, where virtual machines from the reservation marketspaces can be used. The average cost curves for portfolios that do not consider migrations have a local minimum at the planning period of one year. For that planning period, a large share of the optimal portfolios consists of virtual machines that are purchased from the 1-year reservation marketspace. In portfolios where a lower penalty for an interruption of a virtual machine on the spot marketspace is assumed, the share of virtual machines from the 1-year reservation marketspace is lower than in portfolios where a higher penalty is assumed, see figure~\ref{fig:portofliosWithotuMigration}. Therefore, the difference between the average costs for portfolios with a planning period of $1$ year ($8760$ hours) and the following planning period ($1$ year and 1 hour, $8761$ hours) increases with an increasing penalty.  In all three plots, the average costs per hour per virtual machine of portfolios that consider migration have a local maximum at planning periods between $7$ and $13152$ hours.
\begin{itemize}
\item At the planning period of $7$ hours, the cost-efficient 6-hour spot marketspace can not be used without migration. Hence, a migration at hour $6$ is necessary, which requires that a virtual machine be hosted twice at the same time, and so the average costs per hour increase. With a planning period of $10$ hours, the migration costs are distributed over $10$ hours so that the average costs decrease. 

\item A second local maximum occurs at the planning period of $13152$ hours ($1.5$ years). For the previous planning period ($8761$ hours - $1$ year) the cost-effective $1$-year reservation marketspace can be used while for the following planning period ($13152$ - $2$ years)  a virtual machine from 1-year reservation marketspace could be used consecutively. In both cases, the paid fee can be distributed over the complete runtime. For hosting the portfolio for $13152$ hours a combination of the 1-year reservation marketspace and other marketspaces is necessary as figure~\ref{fig:portfolioMigration1000} shows. So the local maximum at the planning period $13152$ hours results from the fact that the contract period of virtual machines from the 1-year reservation marketspace does not perfectly fit, leading to migrations to marketspaces where virtual machines have higher costs.
\end{itemize}

 Especially for planning periods between $7$ hours and $2$ years ($17520$ hours), portfolios that consider migrations have lower costs than portfolios that do not consider migrations. This has two reasons:
\begin{itemize}
\item Virtual machines from the 6-hours spot marketspace have low costs but cannot be used for portfolios with planning periods $>6$ hours if migration is not foreseen. 
\item  Also, virtual machines from the reservation marketspaces are cost-efficient. If the planning period is longer than the contract period of a virtual machine from a reservation marketspace, then consumers have to pay the fee twice if migration is not foreseen. If migration is foreseen, then consumers have greater flexibility as they can use virtual machines from the reservation marketspace and migrate them to alternative marketspaces for the remaining hours of the planning period.
\end{itemize}

\section{Validation}
\label{sec:validation}

In this section, we used the second dataset of the Bitbrains datacenter (RND), which contains $500$ virtual machines that were traced over three months. The virtual machines of this dataset are from a completely different field of application. The previously analyzed dataset contains virtual machines representing application servers, while the virtual machines of the RND dataset were used as management machines. We executed the same analyses for both datasets, but this section focuses on the most important findings.

In a first step, we also executed a mapping process for this dataset to find appropriate instance types on Amazon for the virtual machines of the dataset. 
Thereby, the instance type $t3.nano$ is appropriate for most of the requested virtual machines. Also, a second mapping process was executed where the maximum utilization of the virtual machines, instead of the requested resources, was used. Thereby, 339 (68\%) virtual machines were downgraded to cheaper instance types, which shows that most of the requests are over-sized. So the share of over-sized virtual machines is comparable to the share that we identified for the first dataset. Similar to the first dataset, the Gini index, which measures the concentration of instance types, is lower for the portfolio resulting from the first mapping process ($0.53$) than for the portfolio resulting from the second mapping process ($0.54$).
Choosing instance types based on the maximum utilization instead of the requested resources leads to significant cost-benefits, which are, for example $36.7\%$ on the spot marketspace. These cost savings are comparable to the cost savings that were identified for the first dataset.

\begin{figure}[ht]
\begin{subfigure}[c]{0.6\textwidth}
	\includegraphics[width=0.99\linewidth]{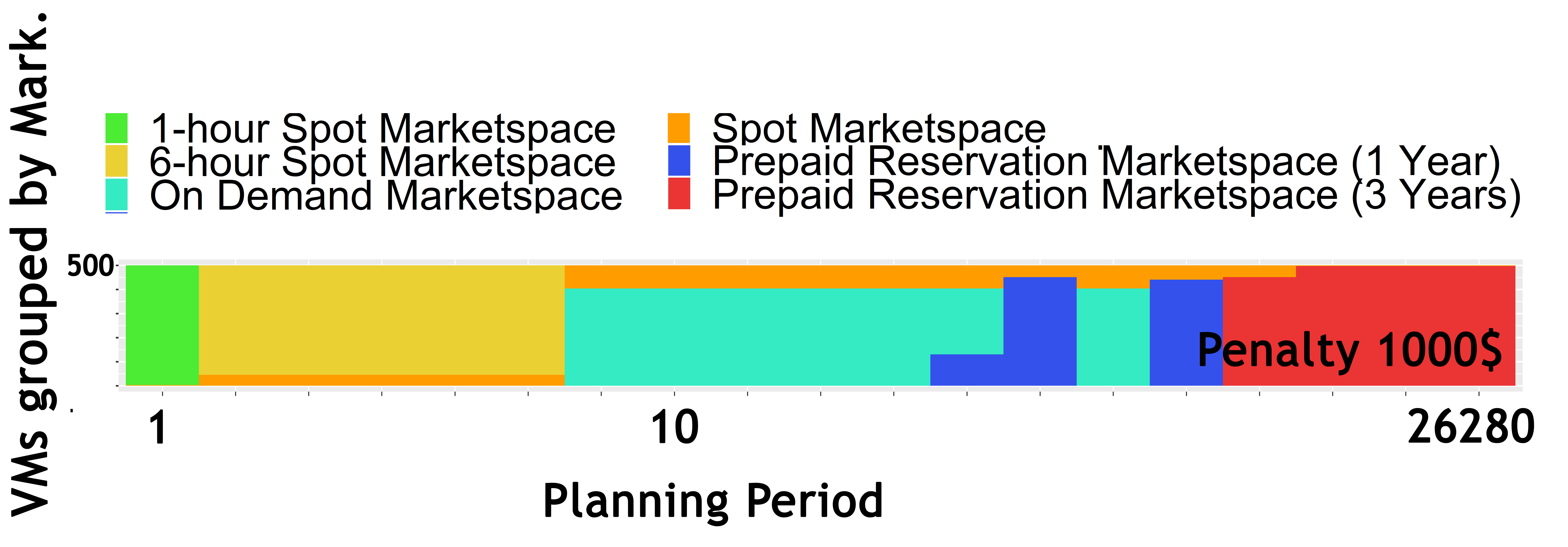}
	\caption{Optimal portfolios}
	\label{fig:validationPortfoliio}
\end{subfigure}
\begin{subfigure}[c]{0.4\textwidth}
	\includegraphics[width=1\linewidth]{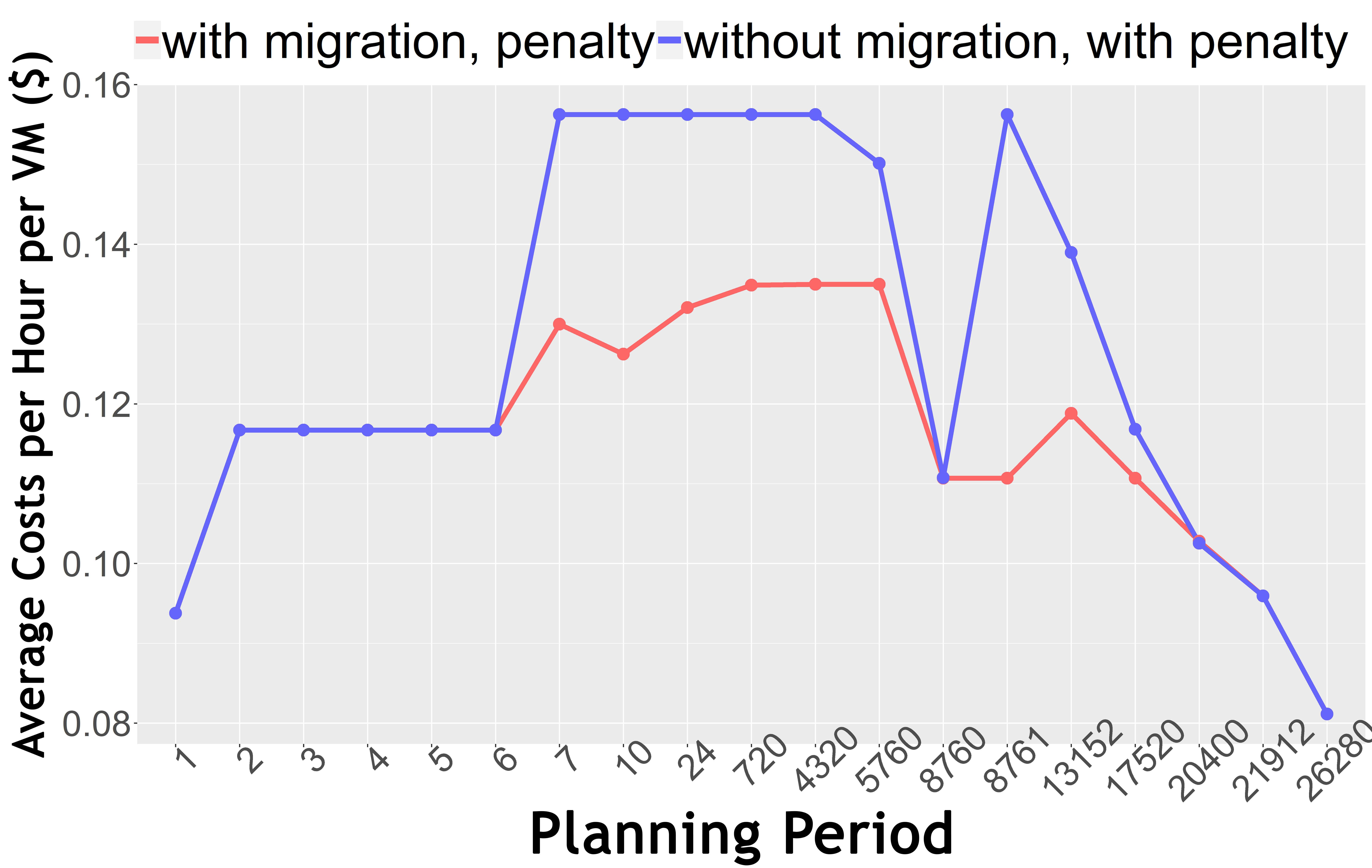}
	\caption{Average costs of a virtual machine per hour of portfolios}
	\label{fig:validationavgCostPortfolio}
\end{subfigure}
\caption{Analysis of RND dataset using a penalty of $1000\$$ without considering migration}
\end{figure}

Exemplary heterogeneous portfolios without migrations are visualized in figure~\ref{fig:validationPortfoliio}. It shows the structure of optimal portfolios assuming a penalty of $1000\$$ for an interruption of a virtual machine on the spot marketspace.  The structure of the portfolios is identical to the structure of the portfolios which were identified for the first dataset: for planning periods up to $6$ hours, most of the virtual machines have to be purchased from the 1 and 6-hour spot marketspace. For longer planning periods, the on-demand marketspace becomes relevant. 
The reservation marketspaces are important to create optimal portfolios with long planning periods. 



Figure~\ref{fig:validationavgCostPortfolio} shows the average costs per hour per virtual machine of the optimal portfolios. The cost structure is comparable to the cost structure that was identified in the dataset, only the height of the costs is different: For long as well as for short planning periods, migrations do not lead to lower costs than portfolios that do not use migrations. However, for portfolios with a planning period between $7$ hours and $2$ years, migrations lead to significant cost savings. 

\section{Conclusion}
\label{sec:conclusion}

Creating optimal cloud portfolios for a given set of requested virtual machines is a critical issue in industry and academia. In the paper at hand, we used a cloud dataset from the Bitbrains datacenter, which contains utilization traces of virtual machines. We treated those virtual machines as requests for virtual machines for which optimal cloud-portfolios were formed using Amazon's marketspaces.  First, appropriate instance types were identified for the requests. Thereby, we found out that most of the requests are oversized - consumers usually do not use the requested resources of virtual machines. This situation has a significant potential for cost optimization. Then, three different types of portfolios were created: homogeneous, heterogeneous without migration, and heterogeneous with migration. The portfolio costs of all three types reach a maximum for planning periods between 7 hours and 3 years. Our analysis shows that heterogeneous portfolios that foresee migrations are most cost-effective. They lead to significant cost savings for planning periods between $7$ hours and $2$ years ($17520$ hours). Thereby, virtual machines from the 6-hour spot marketspace play a vital role: these cost-efficient virtual machines can also be used for portfolios with planning periods exceeding 6 hours if migration is foreseen.   
A second dataset from the Bitbrains datacenter, which contains virtual machines from a different field of application, was used to validate the identified trends and results.

In our further research, we will implement a software tool that acts as a portfolio optimizer. It continuously checks the costs of the marketspaces and executes migrations to achieve cost savings. Further, it will allow for migrations between different availability zones to boost cost savings.



\bibliographystyle{splncs03}
\bibliography{migration}   

\newpage

\section{Online Appendix}
\label{sec:appendix}

\subsection{Gini Index}

The Gini index, which measures the equality of distribution, can be calculated with the following equation:
\begin{equation}
	Gini-index=\sum_{i=1}^{q}{(k_{i-1}+k_{i}) \frac{a_{i}H_{i}}{\sum_{j=1}^{q}{a_{j} H_{j}}}-1}
\end{equation}

$q$ is the number of instance types while $a_{i} H_{i}$ represents the share of virtual machines using the instance type $i$ on the total number of virtual machines. $k_{i}$ represents the share of the instance type on the total number of instance types. Here, an ordering of the instance types is assumed so that for each pair $(k_{i},k_{i+1})$  the condition $k_{i}<k_{i+1}$ holds. After doing the Gini-Correction ($Gini-index_{cor}=\frac{n}{n-1} \cdot Gini-index$) the values can be interpreted as follows: $0$  $\rightarrow$ complete equal distribution which means that all instance types are used for the same number of virtual machines, $1$ $\rightarrow$ complete unequal distribution which means that a single instance type is used for all virtual machines.

\subsection{Average Prices}
\begin{table}[H]

\centering
\scriptsize
\caption{Average costs of a virtual machine per hour in the optimal portfolio}
\begin{tabular}{p{1.1cm}p{1.1cm}p{1.1cm}p{1.1cm}p{1.1cm}p{1.1cm}p{1.1cm}p{1.1cm}p{1.1cm}p{1.1cm}}
\hline
 &    \multicolumn{3}{|c|}{\textbf{Homogeneous Portfolio}} &   \multicolumn{3}{c|}{\textbf{Heterogeneous Portfolio}} &   \multicolumn{3}{c}{\textbf{Heterogeneous Portfolio With Migration}}  \\
\hline
\textbf{Planning Period} &  \textbf{Penalty 100\$} & \textbf{Penalty 500\$} & \textbf{Penalty 1000\$} &  \textbf{Penalty 100\$} & \textbf{Penalty 500\$} & \textbf{Penalty 1000\$}  &  \textbf{Penalty 100\$} & \textbf{Penalty 500\$} & \textbf{Penalty 1000\$}   \\
\hline
1  &  0.0796\$ & 0.1122\$ & 0.1123\$ & 0.0776\$ & 0.1017\$ & 0.1048\$     & 0.0776\$ & 0.1017\$ & 0.1048\$   \\
2  &  0.0796\$ & 0.1451\$ & 0.1451\$ & 0.0781\$ & 0.1178\$ & 0.1295\$     & 0.0781\$ & 0.1178\$ & 0.1295\$   \\
3  &  0.0796\$ & 0.1451\$ & 0.1451\$ & 0.0781\$ & 0.1178\$ & 0.1295\$     & 0.0781\$ & 0.1177\$ & 0.1295\$   \\
4 &   0.0796\$ & 0.1451\$ & 0.1451\$ & 0.0781\$ & 0.1178\$ & 0.1295\$     & 0.0781\$ & 0.1177\$ & 0.1295\$   \\
5  &  0.0796\$ & 0.1451\$ & 0.1451\$ & 0.0781\$ & 0.1178\$ & 0.1295\$     & 0.0781\$ & 0.1177\$ & 0.1295\$   \\
6  &  0.0796\$ & 0.1451\$ & 0.1451\$ & 0.0781\$ & 0.1178\$ & 0.1295\$     & 0.0781\$ & 0.1177\$ & 0.1295\$   \\
7  &  0.0796\$ & 0.1617\$ & 0.2221\$ & 0.0789\$ & 0.1434\$ & 0.1737\$     & 0.0783\$ & 0.1271\$ & 0.1442\$   \\
10  & 0.0796\$ & 0.1617\$ & 0.2221\$ & 0.0789\$ & 0.1434\$ & 0.1737\$     & 0.0782\$ & 0.1243\$ & 0.1401\$   \\
24  & 0.0796\$ & 0.1617\$ & 0.2221\$ & 0.0789\$ & 0.1434\$ & 0.1737\$     & 0.0783\$ & 0.1286\$ & 0.1465\$   \\
720 & 0.0796\$ & 0.1617\$ & 0.2221\$ & 0.0789\$ & 0.1434\$ & 0.1737\$     & 0.0783\$ & 0.1307\$ & 0.1496\$   \\
4320& 0.0796\$ & 0.1617\$ & 0.2221\$ & 0.0789\$ & 0.1434\$ & 0.1737\$     & 0.0783\$ & 0.1306\$ & 0.1497\$   \\
5760 &0.0796\$ & 0.1617\$ & 0.2071\$ & 0.0789\$ & 0.1414\$ & 0.1668\$     & 0.0783\$ & 0.1306\$ & 0.1497\$   \\
8760 &0.0796\$ & 0.1361\$ & 0.1361\$ & 0.0784\$ & 0.1144\$ & 0.1230\$     & 0.0783\$ & 0.1143\$ & 0.1229\$   \\
8761 &0.0796\$ & 0.1617\$ & 0.2221\$ & 0.0789\$ & 0.1434\$ & 0.1737\$     & 0.0783\$ & 0.1143\$ & 0.1229\$   \\
13152&0.0796\$ & 0.1617\$ & 0.1814\$ & 0.0788\$ & 0.1347\$ & 0.1541\$     & 0.0783\$ & 0.1198\$ & 0.1319\$   \\
17520&0.0796\$ & 0.1361\$ & 0.1361\$ & 0.0784\$ & 0.1149\$ & 0.1292\$     & 0.0783\$ & 0.1143\$ & 0.1229\$   \\
20400&0.0796\$ & 0.1227\$ & 0.1227\$ & 0.0782\$ & 0.1060\$ & 0.1138\$     & 0.0782\$ & 0.1060\$ & 0.1141\$   \\
21912&0.0796\$ & 0.1142\$ & 0.1142\$ & 0.0780\$ & 0.1022\$ & 0.1069\$     & 0.0780\$ & 0.1022\$ & 0.1069\$   \\
26280&0.0796\$ & 0.0952\$ & 0.0952\$ & 0.0754\$ & 0.0901\$ & 0.0912\$     & 0.0754\$ & 0.0902\$ & 0.0912\$   \\
\hline
\end{tabular}
\label{tab:averageCosts}
\end{table}
\end{document}